\documentclass[aps,prb,twocolumn,superscriptaddress,longbibliography,endfloat*]{revtex4-1}
\usepackage{graphicx}
\usepackage{color}
\usepackage{amssymb,amsmath}
\usepackage{bm}
\usepackage[english]{babel}
\usepackage{hyperref}
\begin{document}

\title{Imaging the Kondo Insulating Gap on SmB$_6$}

\author{Michael M. Yee}
\thanks{The first two authors contributed equally to this work.}
\author{Yang He}
\thanks{The first two authors contributed equally to this work.}
\affiliation{Department of Physics, Harvard University, Cambridge, MA 02138, USA}
\author{Anjan Soumyanarayanan}
\affiliation{Department of Physics, Harvard University, Cambridge, MA 02138, USA}
\author{Dae-Jeong Kim}
\author{Zachary Fisk}
\affiliation{Department of Physics and Astronomy, University of California, Irvine, California 92697, USA}
\author{Jennifer E. Hoffman}
\email{jhoffman@physics.harvard.edu}
\affiliation{Department of Physics, Harvard University, Cambridge, MA 02138, USA}

\begin{abstract}
\textbf{Topological insulators host spin-polarized surface states which robustly span the band gap and hold promise for novel applications. Recent theoretical predictions have suggested that topologically protected surface states may similarly span the hybridization gap in some strongly correlated heavy fermion materials, particularly SmB$_{\bm{6}}$. However, the process by which the Sm $\bm{4f}$ electrons hybridize with the $\bm{5d}$ electrons on the surface of SmB$_{\bm{6}}$, and the expected Fermi-level gap in the density of states out of which the predicted topological surface states must arise, have not been directly measured. We use scanning tunneling microscopy to conduct the first atomic resolution spectroscopic study of the cleaved surface of SmB$_{\bm{6}}$, and to reveal a robust hybridization gap which universally spans the Fermi level on four distinct surface morphologies despite shifts in the $\bm{f}$ band energy. Using a cotunneling model, we separate the density of states of the hybridized bands from which the predicted topological surface states must be disentangled. On all surfaces we observe residual spectral weight spanning the hybridization gap down to the lowest $T$, which is consistent with a topological surface state.}
\end{abstract}

\maketitle

The classification of solids based on topological invariants has led to the recognition of new electronic phases of matter. The existence of non-trivial topology in band insulators, combined with time reversal or crystal symmetries, gives rise to topologically protected metallic surface states.\cite{FuPRL2007} Potential applications ranging from spintronics to quantum computing have driven intense research efforts into the surface states of topological band insulators such as Bi- and Sn-based chalcogenides.\cite{AndoJPSJ2013, HasanReview2011, QiRMP2011} Recently, it was suggested that similar topological arguments could apply to the more strongly correlated Kondo insulators.\cite{DzeroPRL2010} In these heavy fermion compounds, itinerant electrons screen the local magnetic moments of the lattice in a process known as the Kondo effect.\cite{ColemanHandbook2007} At temperatures below the Kondo coherence temperature ($T^*$) the conduction electrons hybridize with the magnetic moments to open up an energy gap in the density of states (DOS). In a Kondo insulator, the hybridization gap spans the Fermi level $E_F$, causing a metal to insulator transition upon cooling through $T^*$. In a topological Kondo insulator (TKI), protected chiral surface states are predicted to span the Kondo hybridization gap.

Recently, there has been tremendous interest in the heavy fermion material SmB$_6$ as a possible TKI.\cite{DzeroPRL2010, TakimotoJPSJ2011, LuPRL2013} SmB$_6$ undergoes a metal to insulator transition around 50 K\cite{MenthPRL1969, AllenPRB1979, CooleyPRL1995} which has been attributed to hybridization between the $4f$ electrons and $5d$ conduction band. Below $\sim$3 K, a saturation in the resistivity\cite{MenthPRL1969, CooleyPRL1995} indicates a zero temperature conducting channel which could be explained by the existence of topologically protected surface states.\cite{DzeroPRL2010, TakimotoJPSJ2011} This hypothesis has been investigated by a number of recent point contact spectroscopy,\cite{ZhangPRX2013} transport,\cite{KimArxiv2012, WolgastArxiv2012, KimArxiv2013, ThomasArxiv2013} quantum oscillation,\cite{LiArxiv2013} and angle-resolved photoemission spectroscopy (ARPES)\cite{MiyazakiPRB2012, XuArxiv2013, JiangArxiv2013, NeupaneArxiv2013, FrantzeskakisArxiv2013} experiments. The dispersion and orbital chirality of some surface states,\cite{JiangArxiv2013} the half integer Berry phase from Landau levels,\cite{LiArxiv2013} and transport response to magnetic impurities\cite{KimArxiv2013} are strongly suggestive of nontrivial topology in SmB$_6$.

Although evidence is accumulating for topological surface states on SmB$_6$, precise understanding of their properties is presently limited by poor understanding of the hybridization gap within which they emerge. DC transport\cite{MenthPRL1969, CooleyPRL1995, FlachbartPRB2001} and optical reflectivity\cite{TravagliniPRB1984} studies typically report a gap of $\sim$5-10 meV, but both the activation energy fits and the Kramers-Kronig transformations necessary to extract these gap energies may be affected by residual states in the gap.\cite{CooleyPRL1995, FlachbartPRB2001, GorshunovPRB1999} Larger gaps of 19 meV and 36 meV have also been observed by optical transmissivity\cite{GorshunovPRB1999} and Raman spectroscopy,\cite{NyhusPRB1995} respectively. However, transport and optical techniques cannot determine the gap center with respect to $E_F$. Angle-resolved photoemission spectroscopy (ARPES) experiments, which measure filled states only, loosely identify the magnitude of the hybridization gap as the binding energy of the sharp $f$ band just below $E_F$, typically $E_B\sim$14-20 meV.\cite{MiyazakiPRB2012, XuArxiv2013, JiangArxiv2013, NeupaneArxiv2013} However the lack of information on the empty state side makes even the simple question of whether the gap spans the Fermi level elusive.\cite{FrantzeskakisArxiv2013}

Planar tunneling and point contact spectroscopy (PTS/PCS) purport to measure the complete DOS, showing the $T$-dependent opening of a gap $\Delta\sim$14-22 meV.\cite{AmslerPRB1998, FlachbartPRB2001, ZhangPRX2013} However, PCS lineshapes in SmB$_6$-SmB$_6$ junctions vary dramatically with junction size,\cite{FlachbartPRB2001} while PTS and PCS heterojunction experiments have shown an asymmetric peak on the positive energy side of the gap,\cite{AmslerPRB1998, ZhangPRX2013} in contrast to the preponderance of theoretical and experimental evidence for an electron-like conduction band.\cite{LuPRL2013, AlexandrovArxiv2013, MiyazakiPRB2012, XuArxiv2013, JiangArxiv2013, NeupaneArxiv2013, FrantzeskakisArxiv2013} It remains crucial to access the bare DOS and full hybridization gap.

\begin{figure*}
 \includegraphics[width=1.5\columnwidth,clip]{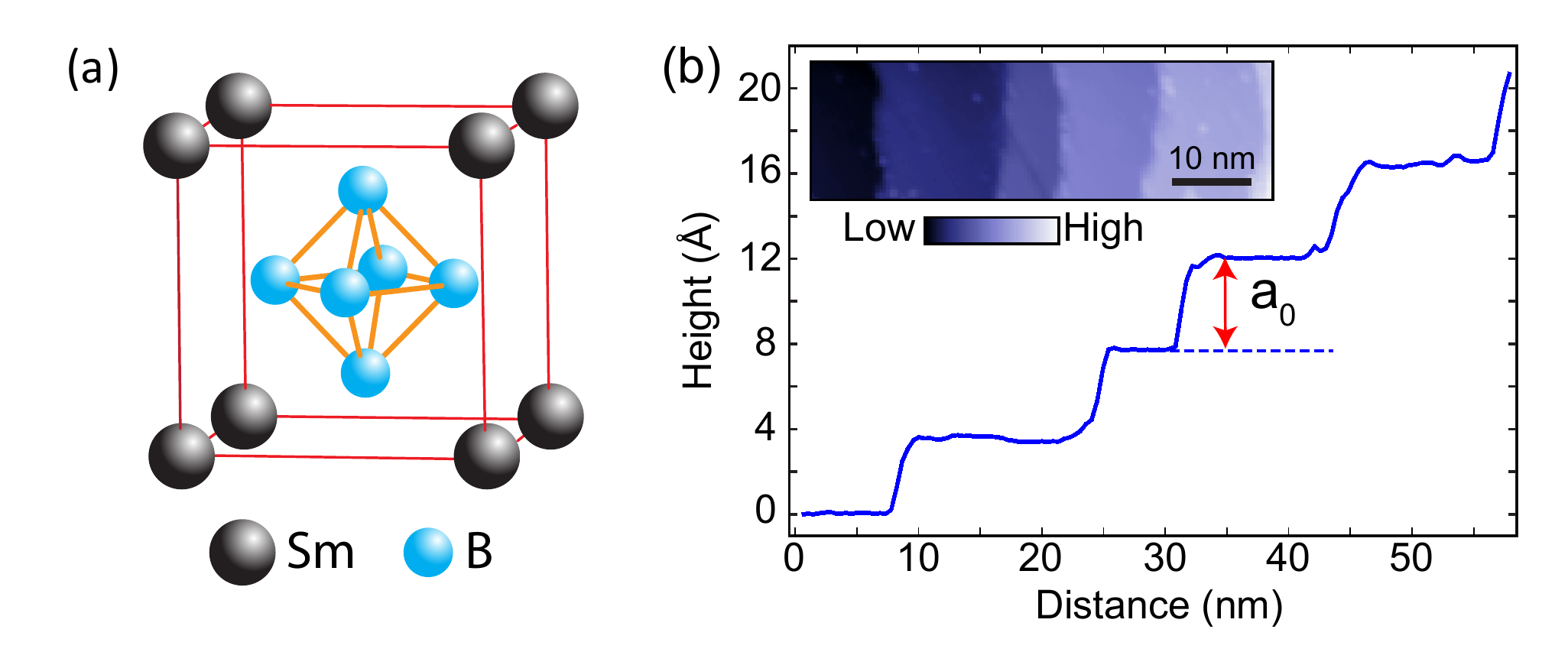}
  \caption{Crystal structure and characterization of SmB$_6$.
  (\textbf{a}) Schematic crystal structure of SmB$_6$ with cubic lattice constant $a_0=4.13$ \AA.
  (\textbf{b}) Topographic linecut across five atomically flat terraces (corresponding to $2\times 1$ morphology in Fig.\ \ref{fig:4morphologies}b). The difference in the vertical height between adjacent terraces is $a_0$.  Inset shows a 50 nm $\times$ 15 nm topography of these terraces. ($T=7$ K; setpoint voltage $V_s=-100$ mV; junction resistance $R_J=10$ G$\Omega$.)
\label{fig:structure}
}
\end{figure*}

The aforementioned studies on SmB$_6$ have averaged over at least several microns of surface area. Spatial averaging over large regions of SmB$_6$ is problematic because, unlike the first generation of Bi-based topological insulators, which are layered materials with natural cleavage planes, SmB$_6$ is a fully three dimensional material whose cleavage properties are unknown. SmB$_6$ has a CsCl-type cubic crystal structure with alternating Sm$^{2+}$ ions and B$_6\,\!^{2-}$ octahedra, shown in Fig.\ \ref{fig:structure}a. It is therefore expected that complete Sm$^{2+}$(001) or B$_6\,\!^{2-}$(001) terminations would be polar, resulting in surface band bending. On the other hand, a partial Sm surface may suffer from structural reconstructions as seen by low energy electron diffraction (LEED).\cite{AonoSurfSci1979, MiyazakiPRB2012} Although the topologically protected surface states are expected to exist on all surface morphologies, their manifestation may be influenced by the differing electronic environments in which they live. Furthermore, the possible shifts of the hybridization gap and/or coexistence of topologically trivial states on some surfaces may short out the fundamental chiral states of interest for transport devices. It remains crucial to quantify the hybridization gap itself, and to understand its variation with surface morphology. Here we use atomically resolved scanning tunneling microscopy and spectroscopy (STM/STS) to probe variations in differential tunneling conductance ($dI/dV$) across multiple SmB$_6$ surface morphologies. We demonstrate that vacuum tunneling conductance is dominated by the bare DOS, and shows a robust hybridization gap which universally spans the Fermi level on all surfaces, as well as anomalous in-gap spectral weight.

\vspace{2mm}
\noindent {\large \textbf{\textsf{Results}}}

\noindent \textbf{Surface structure.} The topographic image in Fig.\ \ref{fig:structure}b shows the cleaved surface of SmB$_6$ with atomically flat terraces of typical $\sim$10 nm extent. Terraces are separated by steps of height equal to the cubic lattice constant $a_0 = 4.13$ \AA, which identifies the cleaved surface as the (001) plane. After the conclusion of the STM experiment, we performed electron back scatter diffraction (EBSD) and x-ray photoelectron spectroscopy (XPS) measurements which confirmed the (001) orientation and showed a B-rich surface,\cite{Supplement} consistent with previous measurements.\cite{AonoSurfSci1979}

\begin{figure*}
 \includegraphics[width=1.9\columnwidth,clip]{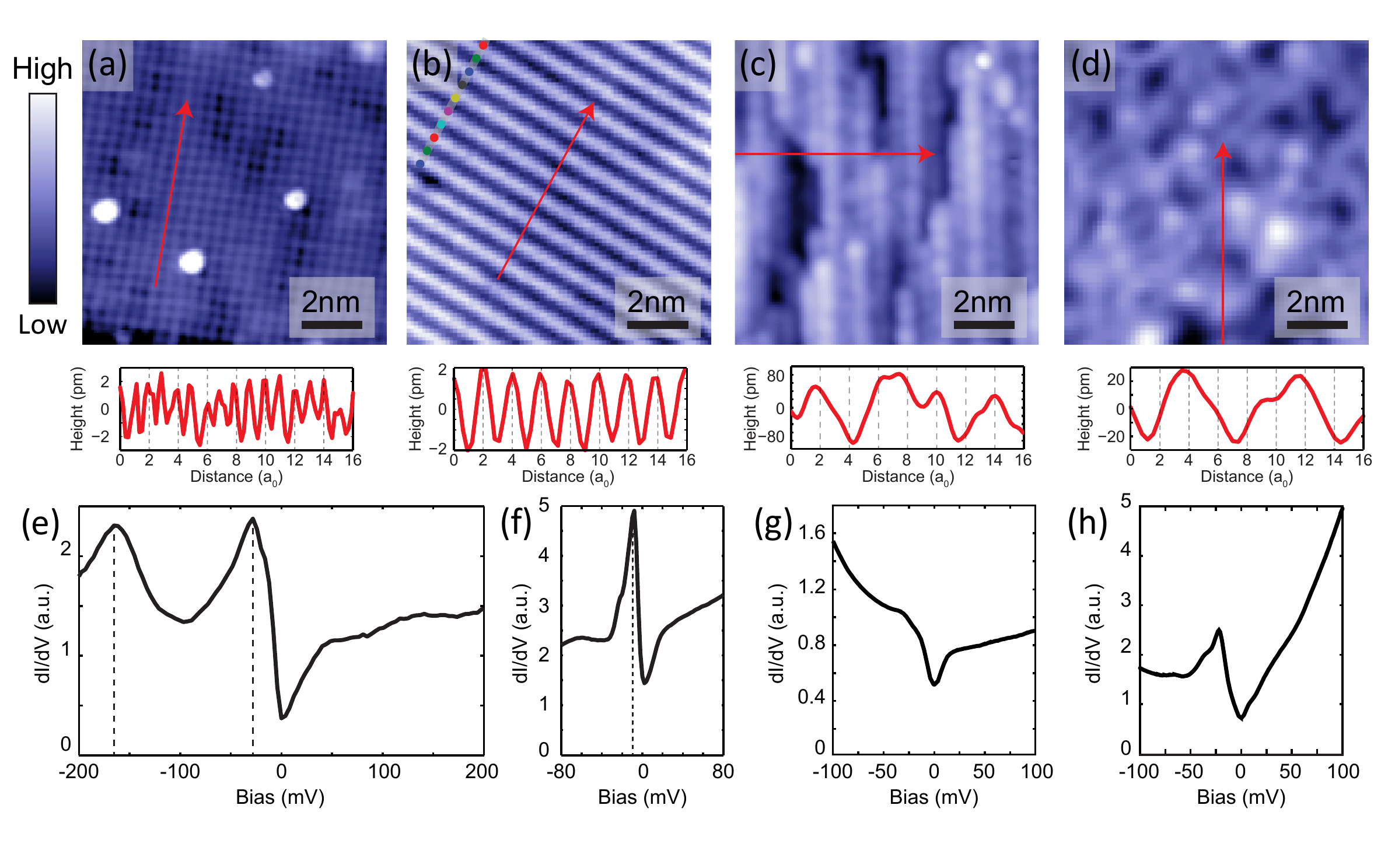}
  \caption{Surface morphology of SmB$_6$ and representative $dI/dV$.
  (\textbf{a}-\textbf{d}) Representative 10 nm $\times$ 10 nm topographic images of the four different surface morphologies, with linecuts along the red arrows showing the surface corrugation beneath each image.
  (\textbf{a}) $1\times 1$ Sm termination. ($T =9.5$ K, $V_s=-200$ mV, $R_J=10$ G$\Omega$.)
  (\textbf{b}) $2\times 1$ half-Sm termination. ($T=8.5$ K, $V_s=-100$ mV, $R_J=5$ G$\Omega$.) Colored dots indicate the locations of spectra displayed in Fig.\ \ref{fig:2by1}a.
  (\textbf{c}) Disordered filamentary B termination. ($T=5.5$ K, $V_s=+200$ mV, $R_J=20$ G$\Omega$.)
  (\textbf{d}) Disordered web termination. ($T=9$ K, $V_s=-100$ mV, $R_J=2$ G$\Omega$.)
  (\textbf{e}-\textbf{h}) Spatially averaged $dI/dV$ representative of each of the four surface morphologies shown in \textbf{a}-\textbf{d}.
  (\textbf{e}) $dI/dV$ on the $1\times 1$ surface. Dashed lines indicate peaks at -165mV and -28mV. ($T=9$ K, $V_s=-250$ mV, $R_J=2$ G$\Omega$, bias excitation amplitude $V_{\mathrm{rms}}=2.8$ mV.)
  (\textbf{f}) $dI/dV$ on the $2\times 1$ surface. Dashed line indicates a peak at -8 mV. ($T=7$ K, $V_s=-100$ mV, $R_J=2$ G$\Omega$, $V_{\mathrm{rms}}=2.1$ mV.)
  (\textbf{g}) Average $dI/dV$ on the disordered filamentary surface. Spectra are very inhomogeneous.\cite{Supplement} ($T=7$ K, $V_s=-150$ mV, $R_J=3$ G$\Omega$, $V_{\mathrm{rms}}=3.5$ mV.)
  (\textbf{h}) Average $dI/dV$ on the disordered web surface. Spectra are very inhomogeneous,\cite{Supplement} with an average a peak at -22 mV.  ($T=9$ K, $V_s=-100$ mV, $R_J=2$ G$\Omega$, $V_{\mathrm{rms}}=2.8$ mV.)
\label{fig:4morphologies}
}
\end{figure*}

Figure 2 shows higher resolution topographies of the four distinct surface morphologies we observed. Figure \ref{fig:4morphologies}a shows a rarely observed $1\times 1$ square lattice which we identify as a complete Sm layer, similar to the complete La layer of (001) cleaved LaB$_6$ previously imaged by STM.\cite{OzcomertSurfSci1992} Because the Sm atoms have a valence of 2+, this polar surface may be energetically unfavorable,\cite{GaoPRB2010} explaining its typical limitation to small regions approximately 10 nm $\times$ 10 nm on the cleaved surface. The polar instability of the $1\times 1$ surface could be resolved by removing half of the Sm atoms from the topmost layer, consistent with the $2\times 1$ striped surface in Fig.\ \ref{fig:4morphologies}b (also shown on the terraces of Fig.\ \ref{fig:structure}b). This surface is consistent with LEED observations of a $2\times 1$ reconstruction\cite{MiyazakiPRB2012} and ARPES observations of band-folding\cite{XuArxiv2013, JiangArxiv2013} on the cleaved SmB$_6$ surface. However, we find that the majority of the cleaved surface is disordered and can be classified as filamentary or amorphous, shown in Figs.\ \ref{fig:4morphologies}c-d, respectively. Both of these disordered surfaces show corrugations $\sim 10 \times$ larger than the suspected Sm terminations in Figs.\ \ref{fig:4morphologies}a-b. Furthermore, the terrace step heights between these disordered morphologies are non-rational multiples of $a_0$. We speculate that the commonly observed filament morphology could be a reconstruction of the B$_6$ octahedra, consistent with our XPS measurements showing that the average surface is B-rich.\cite{Supplement}

\vspace{2mm}
\noindent \textbf{$\bm{f}$ bands and hybridization gap.} Having assigned chemical identities to these surface morphologies, we image their differential tunneling conductance $dI/dV$, which is typically proportional to the local DOS.\cite{BardeenPRL1961} Figs.\ \ref{fig:4morphologies}e-h show spatially averaged spectra representative of each of the four surfaces, emphasizing some ubiquitous features, as well as dramatic differences between the morphologies. The dominant features common to all surfaces are the spectral minimum located near the Fermi energy, and the relative prominence of the peak on the filled state side, compared to the empty state side. Both observations are consistent with the bare DOS for a hybridized electron-like conduction band.\cite{FigginsPRL2010}

To better understand the $f$ band hybridization, we focus in more depth on the two Sm-terminated surfaces. Spectra on the $1\times 1$ surface (Fig.\ \ref{fig:4morphologies}e) show two strong peaks around -165 mV and -28 mV, which we identify as the hybridized Sm$^{2+}$ $^6H_{7/2}$ multiplet typically seen by ARPES at $E_B\sim$150-160 mV, and the hybridized $^6H_{5/2}$ multiplet typically seen by ARPES at $E_B\sim$14-20 meV, respectively.\cite{MiyazakiPRB2012, XuArxiv2013, JiangArxiv2013, NeupaneArxiv2013} The downward energy shift of both the STM-observed $^6H_{7/2}$ and $^6H_{5/2}$ multiplets compared to the average ARPES observations could arise from the polar catastrophe at the $1\times 1$ surface.\cite{NakagawaNatMat2006} The polar catastrophe would cause the movement of electrons towards the surface, decreasing the charge of the surface Sm layer, and would shift the Fermi level up, causing the $f$ bands to appear lower in comparison. Indeed, a very recent ARPES experiment\cite{FrantzeskakisArxiv2013} which boasted no evidence of surface reconstruction from LEED\cite{MiyazakiPRB2012} or band-folding\cite{XuArxiv2013, JiangArxiv2013}, showed similarly higher binding energies of -170 mV and -40 mV, consistent with a chemical potential shift at a polar $1\times 1$ surface.

\begin{figure*}
 \includegraphics[width=1.2\columnwidth,clip]{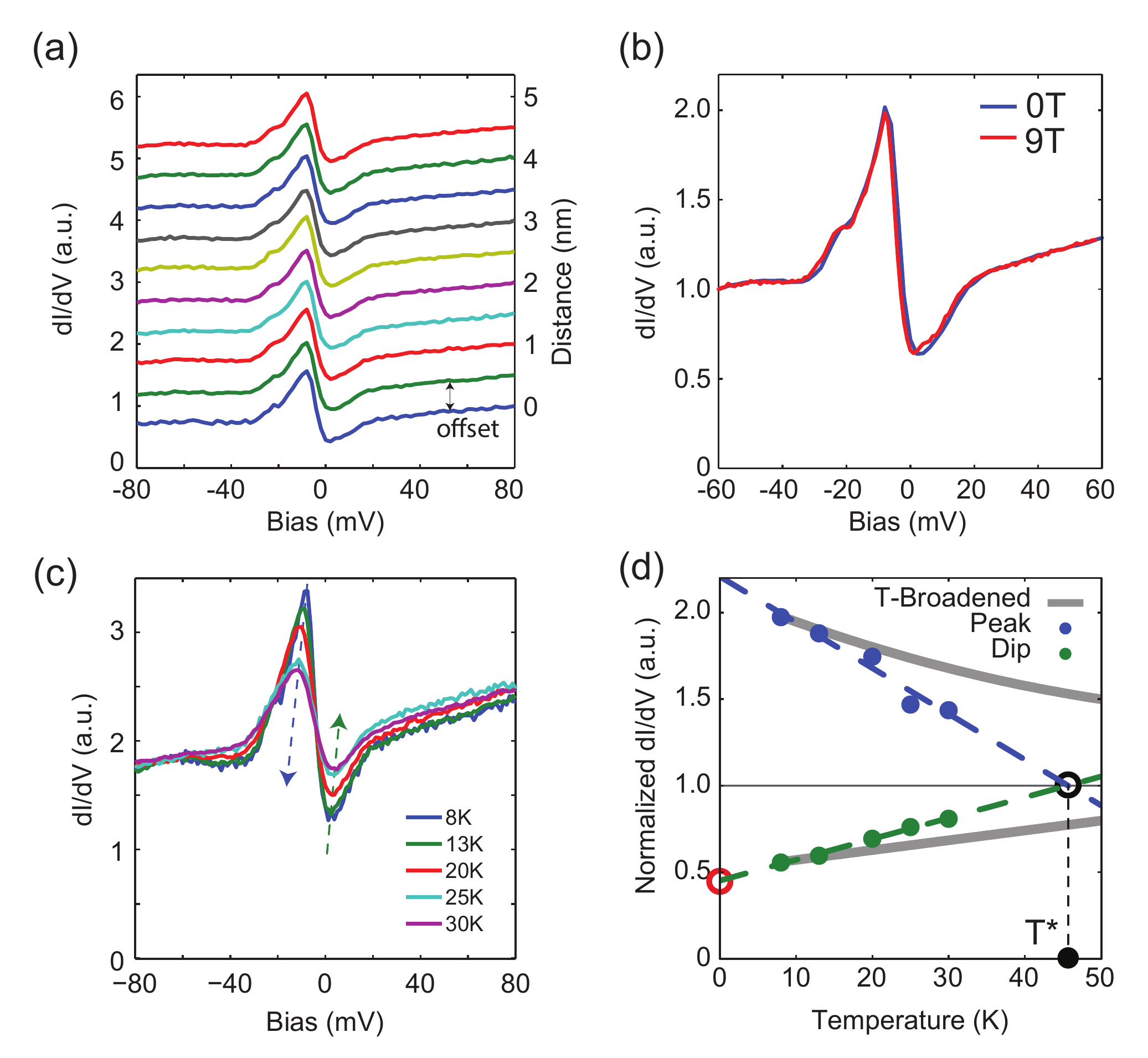}
  \caption{Magnetic field and temperature dependence of spectra on the $2\times 1$ surface.
  (\textbf{a}) $dI/dV$ acquired at the positions of the colored dots in Fig.\ 2b exemplify spectral homogeneity on a clean $2\times 1$ terrace. The spectra have been offset and colored for clarity. ($T=7$ K, $V_s=-100$ mV, $R_J=2$ G$\Omega$, $V_{\mathrm{rms}}=2.1$ mV.)
  (\textbf{b}) Magnetic field independence of the spectra on the $2\times 1$ surface. The -8 meV states shows no change up to 9 T. ($B=0$ T: $T=4.4$ K, $V_s=-100$ mV, $R_J=2$ G$\Omega$, $V_{\mathrm{rms}}=2.1$ mV. $B=9$ T: $T=2.2$ K, $V_s=-60$ mV, $R_J=1.2$ G$\Omega$, $V_{\mathrm{rms}}=1.4$ mV.) (\textbf{c}) Raw $dI/dV$ on clean regions of the $2\times 1$ surface between 8 K and 30 K. ($V_s=-100$ mV, $R_J=1$ G$\Omega$, $V_{\mathrm{rms}}=1.4$ mV.)
  (\textbf{d}) Temperature dependence of the peak (blue) and dip (green) features in normalized $dI/dV$. The simulated reduction (increase) in peak (dip) intensity due to thermal broadening alone are plotted with thick grey lines, and are slower than the corresponding trends in the data. The linear extrapolation of the blue and green data indicate that the peak and dip will vanish around 45 K (black open circle). Extrapolation of the green data to $T=0$ K indicates residual conductance (red open circle) around half of the background conductance.\cite{Supplement}
\label{fig:2by1}
}
\end{figure*}

We expect that the $2\times 1$ surface is nonpolar, and may provide a better view of the bulk $f$ bands and hybridization process. On the $2\times 1$ surface, instead of the broad and inhomogeneous -28 mV feature, we observe a remarkably sharp feature centered at -8 meV (Fig.\ \ref{fig:4morphologies}, which is extremely homogeneous on clean terraces of varying sizes (Fig.\ \ref{fig:2by1}a). The state shows no change in magnetic field up to 9 T (Fig.\ \ref{fig:2by1}b), unlike the field-suppression of the `in-gap' state observed by NMR.\cite{CaldwellPRB2007} A similar state has been seen by some ARPES experiments, weakly dispersing around -8 mV to -4 mV, and has been claimed as the `in-gap' signature of a TKI.\cite{MiyazakiPRB2012, NeupaneArxiv2013} On the contrary, we will argue based on its temperature dependence that the -8 mV feature is the hybridized $4f$-$5d$ band itself, observed specifically on the $2\times 1$ surface and possibly representative of the bulk.

Figure \ref{fig:2by1}c shows substantial reduction in spectral weight of the -8 mV peak between 8 K and 30 K. To determine whether the reduction could be ascribed to thermal broadening alone, we compare the maximum value of the normalized spectra at each temperature to the maximum value of a thermally broadened 8 K spectrum in Fig.\ \ref{fig:2by1}d. It is clear that the spectral peak of the raw data decreases substantially faster than would be expected from thermal broadening alone, and will be completely suppressed by $T\sim$40 - 50 K. We similarly extrapolate that the associated Fermi level gap will be completely filled by $T\sim$40 - 50 K. This temperature scale is consistent with the reported $T^*$ where previous experiments have observed a sharp increase in resistivity,\cite{MenthPRL1969} a sign change of the Hall coefficient,\cite{AllenPRB1979} a change in the magnetic susceptibility,\cite{CaldwellPRB2007} and an abrupt change in Sm valence from 2.50.\cite{MizumakiJPCM2009} This coincident temperature dependence confirms that the -8 mV state is intimately related to the $f$ band hybridization.

\vspace{2mm}
\noindent \textbf{`In-gap' states.} Although the spectral gap width uniformly exceeds 20 meV, we observe residual spectral weight in the gap on all four surfaces. Temperature dependent spectroscopy on the $2\times 1$ surface allows extrapolation of the gap minimum to $T=0$ K to check whether the in-gap spectral weight is consistent with simple thermal excitation. Figure \ref{fig:2by1}d shows that even at zero temperature, the extrapolated minimum $dI/dV$ is nonzero and approximately half the background conductance. This contrasts with a two-channel tunneling model of a clean Kondo insulator in which the hybridization gap should completely suppress the Fermi level tunneling conductance at $T = 0$ K.\cite{MaltsevaPRL2009,FigginsPRL2010} Furthermore, unlike previous STM observations of hybridization gap development in other heavy fermion materials,\cite{SchmidtNature2010, ErnstNature2011, AynajianNature2012} the $2\times 1$ surface of SmB$_6$ measured in Fig.\ \ref{fig:2by1}c is free from quantum critical fluctuations, impurities, or structural defects, which would increase the self-energy and move spectral weight into the gap.\cite{WolflePRL2010} Our observation of zero bias tunneling conductance are consistent with the in-gap surface states of a TKI recently suggested by other experiments,\cite{JiangArxiv2013, LiArxiv2013, KimArxiv2013} but we cannot exclude the possibility of topologically trivial surface states.\cite{FrantzeskakisArxiv2013}

\vspace{2mm}
\noindent {\large \textbf{\textsf{Discussion}}}

It is well known that tunneling into a Kondo impurity -- a single magnetic atom in a non-magnetic host -- reflects the impurity level, the conduction band, {\em and} the quantum mechanical interference between those two tunneling channels. The interference manifests as a Fano resonance -- an asymmetric dip-peak feature -- which dominates the tunneling signal.\cite{MadhavanScience1998} Similarly in Kondo lattice systems, the interference effect may dominate the differential tunneling conductance, giving an asymmetric dip-peak but obscuring the underlying DOS.\cite{YangPRB2009, MaltsevaPRL2009, FigginsPRL2010, WolflePRL2010, BenlagraPRB2011} Here we will separate the three components of the STM-measured $dI/dV$ on SmB$_6$ and show that it is dominated by the bare DOS.

\begin{figure*}
 \includegraphics[width=1.7\columnwidth,clip]{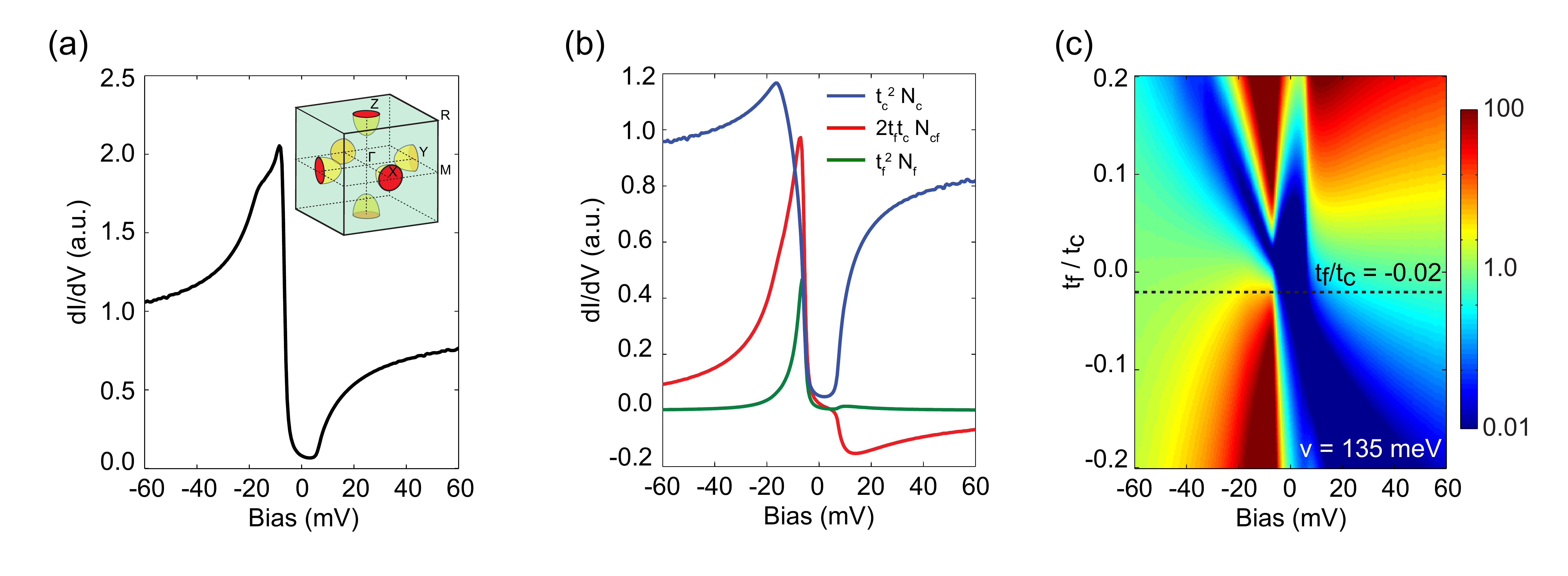}
  \caption{Decomposition of the measured tunneling conductance into DOS and interference channels. (\textbf{a}) Simulation of $dI/dV$ on the $2\times 1$ surface using a two-channel tunneling model.\cite{FigginsPRL2010} The conduction band was modeled as an ellipsoid centered at the $X$ point in the three-dimensional Brillouin zone shown in the inset, and the $f$ band was dispersionless. The main features of the data in Fig.\ \ref{fig:2by1}c are well-matched for self energies $\gamma_c=\gamma_f=0.7$ meV (equivalent to $k_B T$ at the measurement temperature $T=8$ K), $f$ band energy $E_0^f=-4$ meV, hybridization amplitude $v=135$ meV, and tunneling ratio $t_f/t_c=-0.02$.\cite{Supplement} (\textbf{b}) Scaled contributions to $dI/dV$ from the conduction band (blue), $f$ band (green), and interference (red). (\textbf{c}) Simulated $dI/dV$ as a function of $t_f/t_c$ with the other parameters identical to \textbf{a}. The dominant peak position and shape evolve dramatically with $t_f/t_c$; the dashed horizontal line indicates the best match to our data.
\label{fig:fitting}
 }
\end{figure*}

The differential tunneling conductance can be modeled as
\begin{equation}
\frac{dI}{dV}(V)=t_c^2 N_c(V) + t_f^2 N_f(V) + 2 t_c t_f N_{cf}(V)
\end{equation}
\noindent where $N_c(V)$ and $N_f(V)$ are the imaginary parts of the conduction and $f$ band Green's functions, $t_c$ and $t_f$ are the respective tunneling amplitudes, and $N_{cf}(V)$ represents the quantum mechanical interference between the two tunneling channels.\cite{FigginsPRL2010, WolflePRL2010, BenlagraPRB2011, AynajianNature2012} Guided by ARPES,\cite{JiangArxiv2013} we model the $5d$ conduction band as an ellipsoid centered at the $X$ point, shown schematically in the inset to Fig.\ \ref{fig:fitting}a, and the $4f$ states as a flat band whose energy depends on the chemical potential at the surface. We vary the $f$ band energy $E_0^f$, tunneling ratio $t_f/t_c$, and hybridization amplitude $v$ to find the best match to our data, shown in Fig.\ \ref{fig:fitting}a.\cite{Supplement} The fit qualitatively reproduces the peak position, peak shape, and gap width in the data. The scaled components $N_c(V)$, $2(t_f/t_c) N_{cf}(V)$, and $(t_f/t_c)^2 N_f(V)$ are separately plotted in Fig.\ \ref{fig:fitting}b, thus giving access to the bare hybridized DOS in $N_c(V)$ and $N_f(V)$. Both show a dominant peak on the filled state side, in accordance with the electron-like $5d$ conduction band.\cite{MiyazakiPRB2012, XuArxiv2013, JiangArxiv2013, NeupaneArxiv2013, FrantzeskakisArxiv2013} Because the ratio $t_f/t_c$ depends on the details of the tunnel junction, we show in Fig.\ \ref{fig:fitting}c the dependence of $dI/dV$ on $t_f/t_c$. For positive $t_f/t_c$, a prominent peak from the interference term appears on the empty state side, consistent with PTS and PCS data.\cite{AmslerPRB1998, ZhangPRX2013} In fact, given the ARPES-measured bulk band structure, the Green's functions of Fig.\ \ref{fig:fitting}b show that a positive bias peak can come only from $N_{cf}$, suggesting that PTS/PCS experiments are dominated by the interference rather than the bare DOS. The dip apparent in PTS/PCS data may therefore represent an energy range of destructive interference, and not necessarily the hybridization gap. But for all four surface morphologies in Figs.\ 2e-h, we find the more prominent peak on the filled state side of the spectral gap, consistent with expectations for the bare DOS, suggesting that the STM-observed spectral gap is representative of the true hybridization gap. Furthermore, throughout this $t_f/t_c < 0$ tunneling regime, the modeled differential conductance is vanishingly small within the hybridization gap, in contrast to the in-gap spectral weight we observe on all four surfaces.

We use the spatial resolution of STM to reconcile some apparent discrepancies between ARPES experiments.\cite{MiyazakiPRB2012, XuArxiv2013, JiangArxiv2013, NeupaneArxiv2013, FrantzeskakisArxiv2013} ARPES, with typical spot size on the order of hundreds of microns, measures signatures of all surfaces simultaneously. We expect that most ARPES experiments will show momentum-resolved contributions from both the Sm-terminated $1\times 1$ and $2\times 1$ surfaces (Figs.\ 2a-b), and possibly the top few layers of the bulk, but not from the two disordered surfaces where $k$ is a poor quantum number (Figs.\ 2c-d). Depending on the fractional composition of the cleaved surface structure, as well as the photon energy, depth probed, and detector resolution, ARPES may observe the spatial average of the -28 mV and -8 mV $f$ bands from the Sm-terminated surfaces
as a single broadened $f$ band at intermediate energy,\cite{XuArxiv2013, JiangArxiv2013} or as one\cite{FrantzeskakisArxiv2013} or two\cite{MiyazakiPRB2012, NeupaneArxiv2013} separate states. In the latter scenario, the -8 mV $f$ band has been interpreted as an `in-gap' state.\cite{MiyazakiPRB2012, NeupaneArxiv2013} However, we note that a topological in-gap state would be expected to span the upper and lower hybridized bands, and thus would appear as continuous spectral weight filling the hybridization gap, rather than as a sharp peak at a specific energy. Indeed, we consistently observe broad in-gap spectral weight on all surfaces.

With these first atomically resolved spectroscopic measurements on SmB$_6$, we provide a general new paradigm for interpreting Kondo hybridization, and lay the specific groundwork for understanding TKIs. First, our explicit decomposition of the measured tunneling conductance into DOS vs.\ interference channels provides an intuitive way to understand tunneling measurements of Kondo hybridization in a broad class of heavy fermion materials.\cite{SchmidtNature2010, ErnstNature2011, ParkPRL2012, AynajianNature2012, ZhangPRX2013} Second, we confirm that SmB$_6$ is a Kondo insulator, by using this decomposition to reveal the full $f$ band hybridization gap, spanning the Fermi level, on all four observed surface morphologies. Our temperature dependent spectroscopy illustrates the surface hybridization process, beginning around $T^*\sim$40-50 K, in agreement with previous bulk measurements. Third, our observation of the $f$ band shifts between polar and non-polar cleaved surfaces of SmB$_6$ reveals the dramatically different electronic environments in which the predicted topological surface state must exist. However, we observe in-gap spectral weight significantly exceeding bulk hybridization models persisting on all surface morphologies. Theoretical modeling of bulk band shifts, surface states, and hybridization for different surface terminations is urgently needed. Our work provides the nanoscale spectroscopic details necessary for understanding the first strongly correlated topological insulator.

\vspace{2mm}
\noindent {\large \textbf{\textsf{Methods}}}

\noindent Single crystals of SmB$_6$ were grown using an Al flux method,\cite{KimArxiv2012} cleaved in cryogenic ultrahigh vacuum around 30 K, and immediately inserted into our homebuilt STM. STM tips were cut from PtIr wire and cleaned via field emission on polycrystalline Au foil. We imaged two samples, with multiple tip-sample approaches on each cleaved surface, in regions separated by many microns. Spectroscopic measurements were carried out between 2 and 30 K, in fields up to 9 T, using a standard lock-in technique with bias modulation at 1115 Hz.

\vspace{2mm}
\noindent {\large \textbf{\textsf{Acknowledgements}}}

\noindent We especially thank Piers Coleman, Ilya Elfimov, Rebecca Flint, Victor Galitski, Laura Greene, Dirk Morr, Wan Kyu Park, and Jason Zhu for helpful conversations. We thank Dennis Huang, Eric Hudson, and Can-Li Song for careful reading of the manuscript.
We thank David Lange for help with the SEM and EBSD measurements, and Greg Lin for help with the XPS measurements. M.M.Y. acknowledges a fellowship from NSERC. The work at Harvard was supported by the US National Science Foundation under grant DMR-1106023. The work at UC Irvine was supported by NSF-DMR-0801253 and UC Irvine CORCL Grant MIIG-2011-12-8.

\vspace{2mm}
\phantom{{\Huge O}}
\noindent {\large \textbf{\textsf{References \phantom{{\Huge O}}}}}
\vspace{-1.7cm}

%


\clearpage
\onecolumngrid

{\Large \textbf{Supplemental Material for:}}
\begin{center}
{\large \textbf{Imaging the Kondo Insulating Gap on SmB$_6$}}

Michael M.\ Yee, Yang He, Anjan Soumyanarayanan, Dae-Jeong Kim, Zachary Fisk, Jennifer E.\ Hoffman
\end{center}

\setcounter{figure}{0}
\setcounter{equation}{0}
\setcounter{table}{0}
\makeatletter
\renewcommand{\thefigure}{S\@arabic\c@figure}
\renewcommand{\theequation}{S\@arabic\c@equation}
\renewcommand{\thetable}{S\@arabic\c@table}

\vspace{2mm}
\noindent {\large \textbf{\textsf{I. Surface characterization}}}

\begin{figure*}[!h]
 \includegraphics[width=0.75\columnwidth,clip]{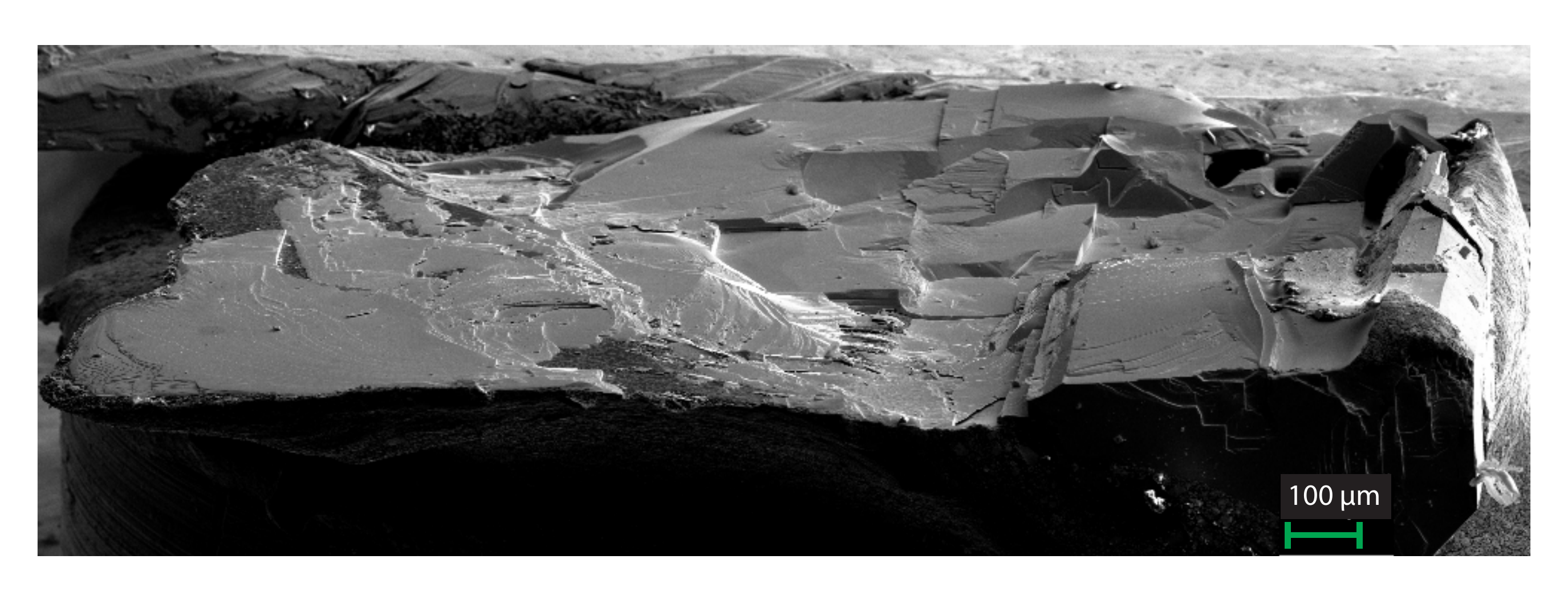}
  \caption{Scanning electron microscope (SEM) image of one of the two SmB$_6$ crystals studied. The crystal was cleaved in cryogenic ultra-high vacuum (UHV) at $T=30$ K and immediately inserted into the STM. After STM experiments were completed, the crystal was removed from vacuum and the cleaved surface was imaged at room temperature at the Center for Nanoscale Systems, Harvard University. All four nanoscale morphologies were observed on this cleaved surface.
\label{fig:SEM}
 }
\end{figure*}

\begin{figure*}[!h]
 \includegraphics[width=0.65\columnwidth,clip]{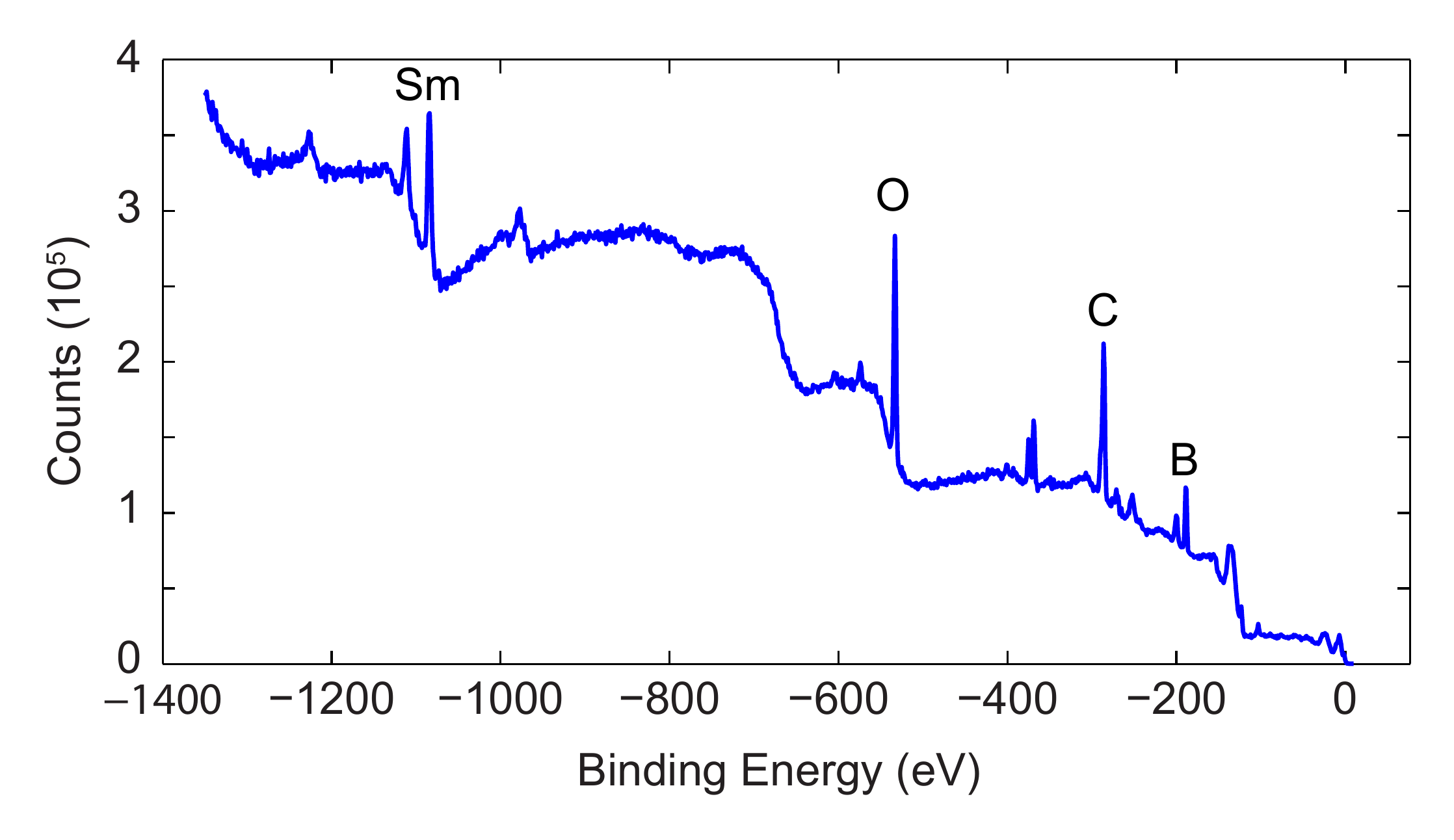}
  \caption{X-ray photoemission spectroscopy (XPS) spectrum of the surface shown in Fig.\ \ref{fig:SEM}. The XPS spectrum was taken after STM measurements were completed at room temperature at the Center for Nanoscale Systems, Harvard University. Annotated peaks correspond to the binding energies of the respective atoms.
\label{fig:XPS}
 }
\end{figure*}

\begin{table*}[!h]
\caption{\label{tab:XPScomp}Atomic composition extracted from the XPS spectrum of Fig.\ \ref{fig:XPS}. The surface ratio of Sm:B is 1:9.5.}
\begin{tabular}{l r@{}l}
    \hline
    Element & \multicolumn{2}{c}{Atomic \%} \\
    \hline\hline
    O $1s$ & 21 & .03  \\
    Sm $3d^5$ & 2 & .56 \\
    C $1s$ & 45 & \\
    B $1s$ & 24 & .34 \\
    Other & 7 & .06 \\
    \hline
  \end{tabular}
\end{table*}

\vspace{2mm}
\noindent {\large \textbf{\textsf{II. Spatial dependence of spectra}}}

\begin{figure*}[!h]
 \includegraphics[width=1\columnwidth,clip]{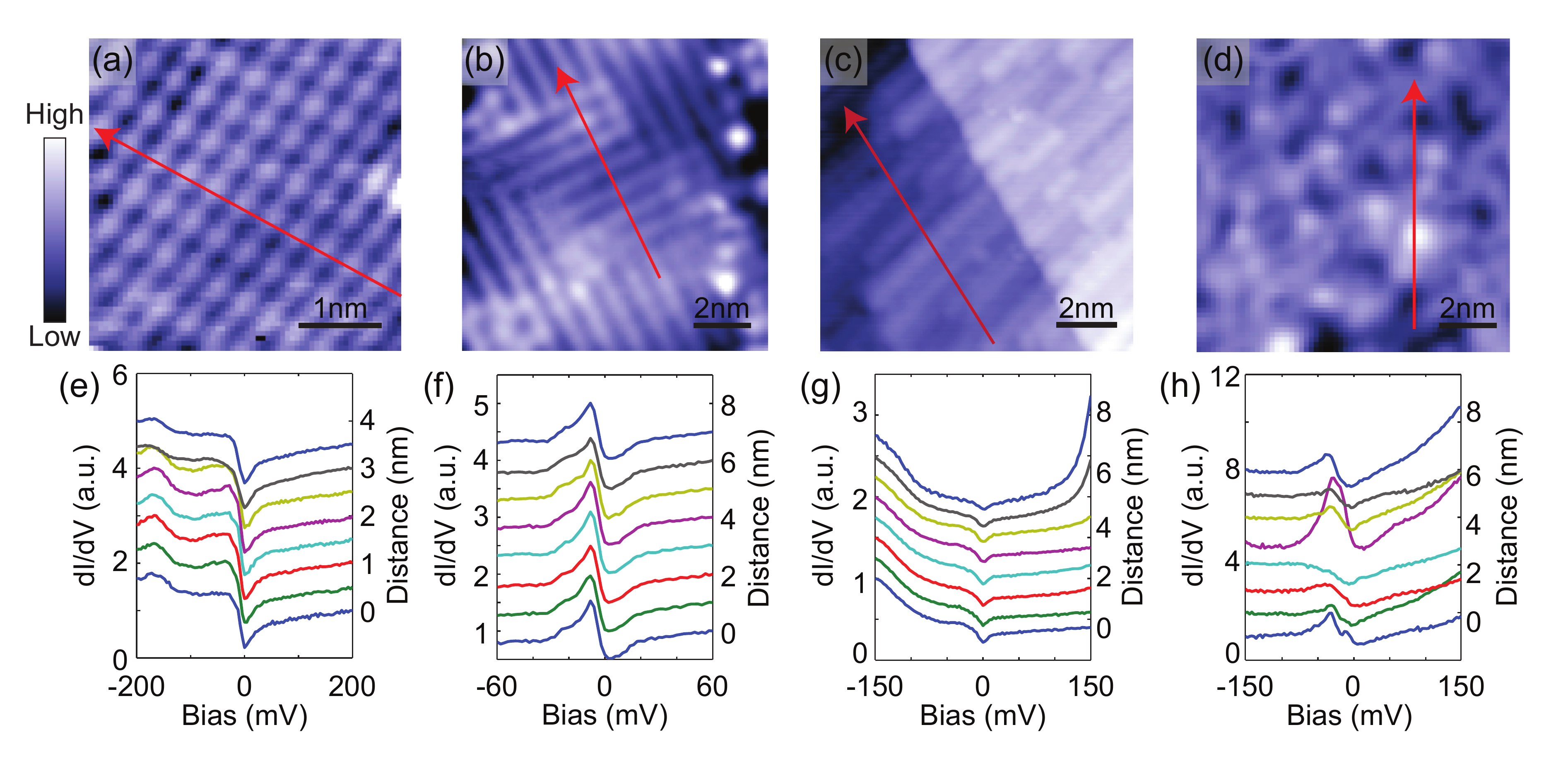}
  \caption{Spatial dependence of differential tunneling conductance. (\textbf{a}-\textbf{d}) Simultaneously acquired topographies from spectral maps on four different surface morphologies, respectively. (\textbf{e}-\textbf{h}) $dI/dV$ spectroscopy along the trajectories denoted by red arrows in panels \textbf{a}-\textbf{d}. (\textbf{a}, \textbf{e}) The $1\times 1$ Sm-terminated region shows weak spectral inhomogeneity ($T = 9.5$ K, $V_s = -200$ mV, $R_J = 2$ G$\Omega$, $V_{\mathrm{rms}} = 2.8$ mV). (\textbf{b}, \textbf{f}) The $2\times 1$ Sm-terminated region shows extreme spectral homogeneity on multiple terraces, despite twin boundaries and phase slips in the reconstruction ($T = 4.4$ K, $V_s = -100$ mV, $R_J = 2$ G$\Omega$, $V_{\mathrm{rms}} = 2.1$ mV). (\textbf{c}, \textbf{g}) The filamentary B-terminated region is spectrally inhomogeneous ($T = 7$ K, $V_s = -150$ mV, $R_J = 3$ G$\Omega$, $V_{\mathrm{rms}} = 3.5$ mV). (\textbf{d}, \textbf{h}) The disordered web region shows extreme spectral inhomogeneity ($T = 9$ K, $V_s = -100$ mV, $R_J = 2$ G$\Omega$, $V_{\mathrm{rms}} = 2.8$ mV).
\label{fig:inhomogeneity}
 }
\end{figure*}

\vspace{2mm}
\noindent {\large \textbf{\textsf{III. Normalization \& thermal broadening}}}

\begin{figure*}[!h]
 \includegraphics[width=0.9\columnwidth,clip]{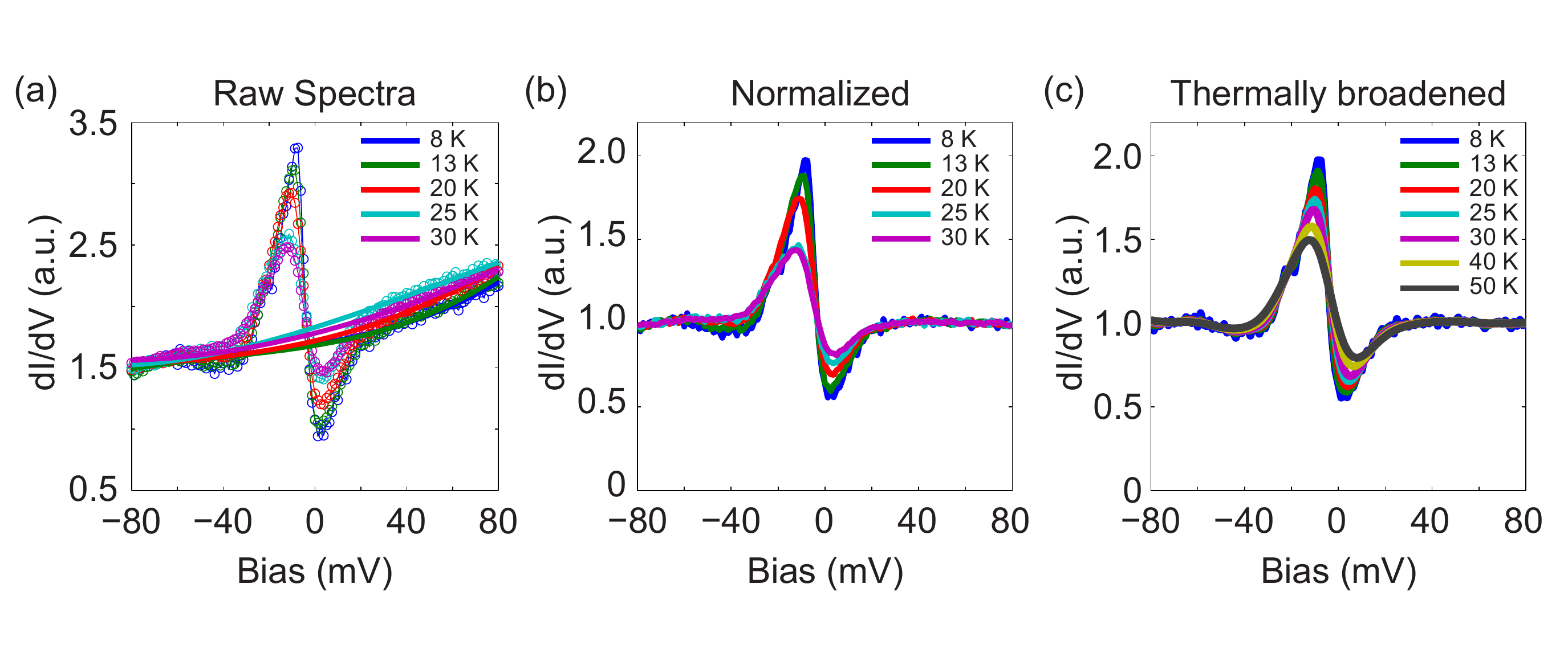}
  \caption{Normalization and simulated thermal broadening. (\textbf{a}) Raw temperature dependent STM spectra (dots, thin line) and fitted third order polynomial backgrounds (thick lines). The polynomial backgrounds were fit to the spectrum excluding the energy range -60 mV $< V < +20$ mV. ($V_s = -100$ mV, $R_J = 1$ G$\Omega$, $V_{\mathrm{rms}} = 1.4$ mV) (\textbf{b}) Normalized spectra obtained by dividing the raw spectra by the fitted polynomial backgrounds of \textbf{a}. The maxima and minima of these normalized spectra are plotted as blue and green points, respectively, in Fig.\ 3d of the main text. (\textbf{c}) Thermally broadened normalized spectra. The $T=8$ K normalized spectrum from \textbf{b} was deconvolved to $T=0$ K then thermally broadened to the simulated temperature labeled in \textbf{c}. Maxima and minima of this sequence of spectra are plotted as grey lines in Fig.\ 3d of the main text.
\label{fig:thermal}
 }
\end{figure*}

Low temperature differential conductance spectra on the $2\times 1$ surface of SmB$_6$ show a prominent peak-dip feature that decreases with increasing temperature (main text Fig.\ 3c). To determine whether the peak-dip reduction is consistent with thermal broadening alone, we first normalize the spectra to remove any artifact variations in asymmetry from $z$ piezo drift, and then simulate the effect of thermal broadening on the normalized spectra.
To eliminate variations in spectral asymmetry arising from $z$ piezo drift we divide each $dI/dV$ spectrum by a cubic polynomial fit to the data, excluding the energy range of the strongly $T$-dependent peak-dip feature (-60 mV $< V < 20$ mV), as shown in Fig.\ \ref{fig:thermal}a. The resultant normalized spectra are shown in Fig.\ \ref{fig:thermal}b.

To study the intrinsic temperature dependence of the $dI/dV$ spectra we need to account for the thermal broadening of data acquired at different $T$. The effect on the $dI/dV$ spectra can be expressed as the convolution of the sample density of states and the derivative of the Fermi-Dirac distribution,\cite{NagaokaPRL2002}
\begin{equation}
\frac{dI}{dV} (V,T)= \int \rho_t(T) \rho_s(E,T) \frac{d}{dV} F(E-eV,T)dE \propto \rho_s(V,T) * \frac{d}{dV} F(V,T)
\end{equation}
\noindent Here $\rho_t$ and $\rho_s$ are the density of states of the tip and sample respectively, * represents a convolution, and the derivative of the Fermi-Dirac distribution is
\begin{equation}
\frac{d}{dE} F(E,T) = \frac{-1}{4k_B T \cosh^2(E/2k_B T)}
\end{equation}
\noindent which has a FWHM of $\sim3.5 k_B T$. Using this formalism we can change the temperature of a spectrum from the temperature at which the data was acquired, $T_{\mathrm{data}}$, by deconvolving the spectrum to $T_0=0$ K with $\frac{d}{dE} F(E,T_{\mathrm{data}})$, and then convolving the spectrum to an arbitrary simulation temperature $T_{\mathrm{sim}}$. We apply this technique to the normalized $T_{\mathrm{data}} = 8$ K spectrum in Fig.\ \ref{fig:thermal}b to arrive at the series of simulated spectra at $T_{\mathrm{sim}}$ in Fig.\ \ref{fig:thermal}c.
\\

\vspace{5mm}
\noindent {\large \textbf{\textsf{IV. Simulating the $\bm{dI/dV}$ spectra}}}

In typical STM/STS experiments, the measured differential conductance $dI/dV$ is representative of the sample density of states (DOS). However, in Kondo systems the measured $dI/dV$ represents the tunneling into the conduction band and the heavy band, as well as the quantum mechanical interference between these two tunneling channels. It is necessary to account for this cotunneling process in order to extract the underlying DOS of the hybridized bands from the $dI/dV$ spectra. We used three models to simulate our experimental spectra: a Fano model\cite{YangPRB2009, WolflePRL2010} as well as two Kondo lattice models.\cite{MaltsevaPRL2009, FigginsPRL2010} While the Fano model fails to fit the SmB$_6$ spectra, both Kondo lattice models capture the main qualitative features of the spectra: the peak location, width, and sharpness, as well as the width and positive energy kink of the gap.

The tunneling conductance into a single ion Kondo system follows a simple Fano lineshape,\cite{MadhavanScience1998}
\begin{equation}
\frac{dI}{dV}(V) \propto \frac{(q+\epsilon)^2}{1+\epsilon^2}
\end{equation}
\noindent where $q$ is the Fano parameter, $\epsilon=(eV-E_0^f)/w$, $E_0^f$ is the energy of the discrete $f$-state and $w$ is the width of the resonance. The Fano lineshape is also the limiting case in Kondo lattice systems with spatial disorder or large self-energy.\cite{WolflePRL2010}

To expand this to a clean Kondo Lattice system, we use both the analytic model of Maltseva, Dzero, and Coleman\cite{MaltsevaPRL2009} and the numerical model of Figgins and Morr.\cite{FigginsPRL2010}

Using the formalism of Figgins and Morr, we modeled the $dI/dV$ spectrum as the sum of three terms from the conduction band, the $f$-band and the interference of the two channels:
\begin{equation}
dI/dV \propto N_c + \left(\frac{t_f}{t_c} \right)^2 N_f +2\left(\frac{t_f}{t_c}\right) N_{cf}
\end{equation}
\begin{equation}
N_c=\mathrm{Im}\left[G_c(k,\omega)\right];\quad
N_f=\mathrm{Im}\left[G_f(k,\omega)\right];\quad
N_{cf}=\mathrm{Im}\left[G_{cf}(k,\omega)\right]
\end{equation}
\noindent where $N$ represents the DOS of the respective channel, $t_f/t_c$ is the ratio of the tunneling amplitudes into the $f$-band and the conduction band, and the hybridized Green's functions are given by
\begin{eqnarray}
G_c(k,\omega) & = & \left[ G_c^0(k,\omega)^{-1} - v^2 G_f^0(k,\omega) \right]^{-1} \\
G_f(k,\omega) & = & \left[ G_f^0(k,\omega)^{-1} - v^2 G_c^0(k,\omega) \right]^{-1} \\
G_{cf}(k,\omega) & = & G_c^0(k,\omega) v G_f(k,\omega) .
\end{eqnarray}
\noindent The hybridized Green's functions are expressed in terms of the hybridization amplitude $v$ and the bare Green's functions: $G_c^0 (k,\omega)= \left[\omega+i\gamma-E_k^c \right]^{-1}$ and $G_f^0 (k,\omega)=\left[\omega+i\gamma-E_k^f \right]^{-1}$ where $E_k^c$ and $E_k^f$ are the unhybridized band structures of the conduction and $f$-band, respectively. The hybridized bands take the form
\begin{equation}
E_k^{\pm}=\frac{1}{2} \left(E_k^c  + E_k^f \right) \pm \sqrt{\frac{1}{2} \left(E_k^c- E_k^f \right)^2+v^2 } .
\end{equation}

We modeled the Sm $5d$ conduction band as an ellipsoid centered at the $X$ point of the three-dimensional Brillouin zone, with semi-major $k_F$ axes $0.401(\pi/a_0) \times 0.401(\pi/a_0) \times 0.600(\pi/a_0)$ and $E_{\mathrm{min}}=-1.6$ eV in agreement with ARPES measurements.\cite{JiangArxiv2013} We modeled the Sm $4f$ band as a non-dispersive flat band spanning the Brillouin zone at energy $E_0^f$. We used a self-energy $\gamma=k_B T$ for the measurement temperature $T=8$ K , and varied $E_0^f$, $t_f/t_c$, and $v$ to match the data. We found a good match to the main qualitative features of the data for $E_0^f=-4$ meV, $v=135$ meV, and $t_f/t_c = -0.02$.

To verify our results we also used the model of Maltseva, Dzero and Coleman:
\begin{equation}
\frac{dI}{dV}(V) \propto \mathrm{Im} \left[ G^{KL}(eV-i\gamma) \right]
\end{equation}
\begin{equation}
G^{KL}(eV)=\left( 1+\frac{q w}{eV-E_0^f} \right)^2  \ln \left[\frac{eV+D_1-\frac{v^2}{eV-E_0^f}}{eV-D_2-\frac{v^2}{eV-E_0^f}} \right] +\frac{D_1+D_2}{eV-E_0^f} \left( \frac{t_f}{t_c} \right)^2
\end{equation}
\noindent where $q=(v/w)(t_f/t_c)$, $-D_1$ and $D_2$ are the lower and upper conduction band edges, $w$ is the width of the single ion Kondo resonance, $v$ is the hybridization amplitude, $t_f/t_c$ is the ratio of the tunneling amplitudes, $E_0^f$ is the energy of the non-dispersive $f$-band and $\gamma$ is the self-energy.\cite{MaltsevaTypo}

We used experimental data on SmB$_6$ to choose appropriate values for the parameters. We used $D_1=1.6$ eV in agreement with ARPES,\cite{JiangArxiv2013} $D_2=3$ eV in agreement with LDA calculations,\cite{AntonovPRB2002} giving a total conduction bandwidth of $D=4.6$ eV.\cite{UnphysicalBandwidth} We used $w=k_B T$ with $T=100$ K in agreement with the onset of a valence change.\cite{MizumakiJPCM2009} We used $\gamma=k_B T$ for the measurement temperature $T=8$ K and varied $E_0^f$, $t_f/t_c$, and $v$ to match the data. We also found good match to the main qualitative features of the data for $E_0^f=-2$ meV, $v=135$ meV, and $t_f/t_c = -0.02$. Furthermore, this value of $v$ is roughly consistent with the expected magnitude of the hybridization gap $\Delta_g \sim 2v^2/D$.

\begin{figure*}[h!]
 \includegraphics[width=0.8\columnwidth,clip]{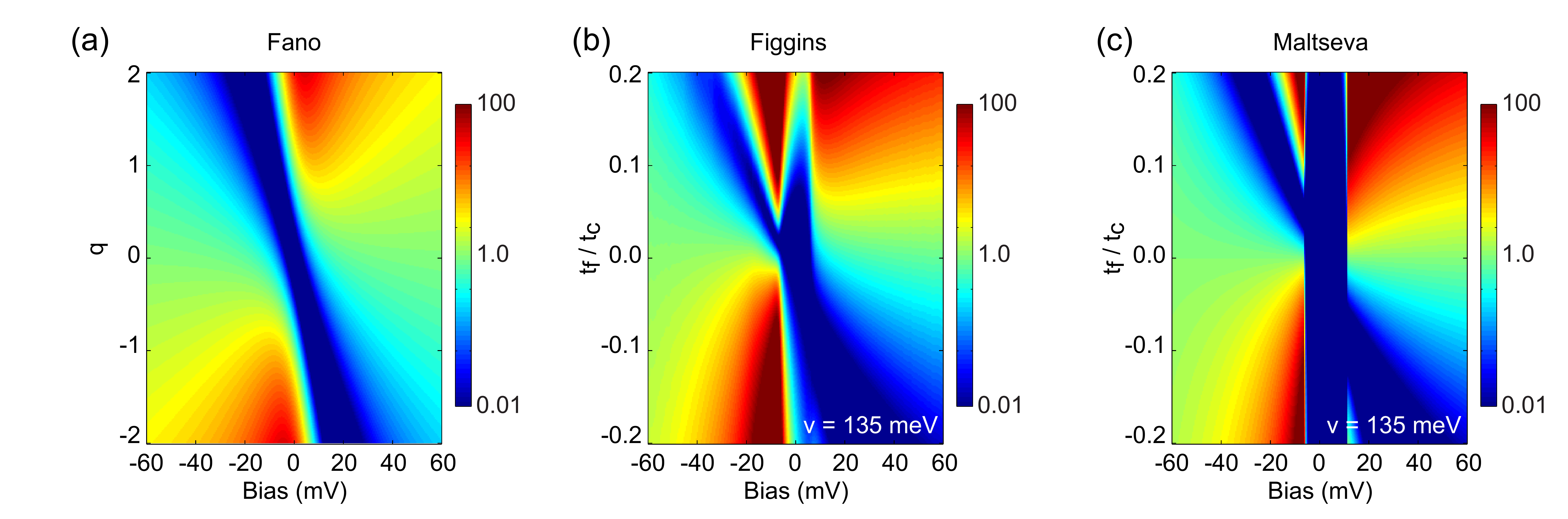}
  \caption{$dI/dV$ of the Fano, Figgins and Maltseva models.
	(\textbf{a}) $dI/dV$ calculated using the Fano lineshape with $w=8.6$ meV.
	(\textbf{b}) $dI/dV$ calculated using the Figgins and Morr model with $v=135$ meV.
	(\textbf{c}) $dI/dV$ calculated using the model of Maltseva, Dzero and Coleman with $v=135$ meV.
\label{fig:varytf}
 }
\end{figure*}

\begin{figure*}[h!]
 \includegraphics[width=0.8\columnwidth,clip]{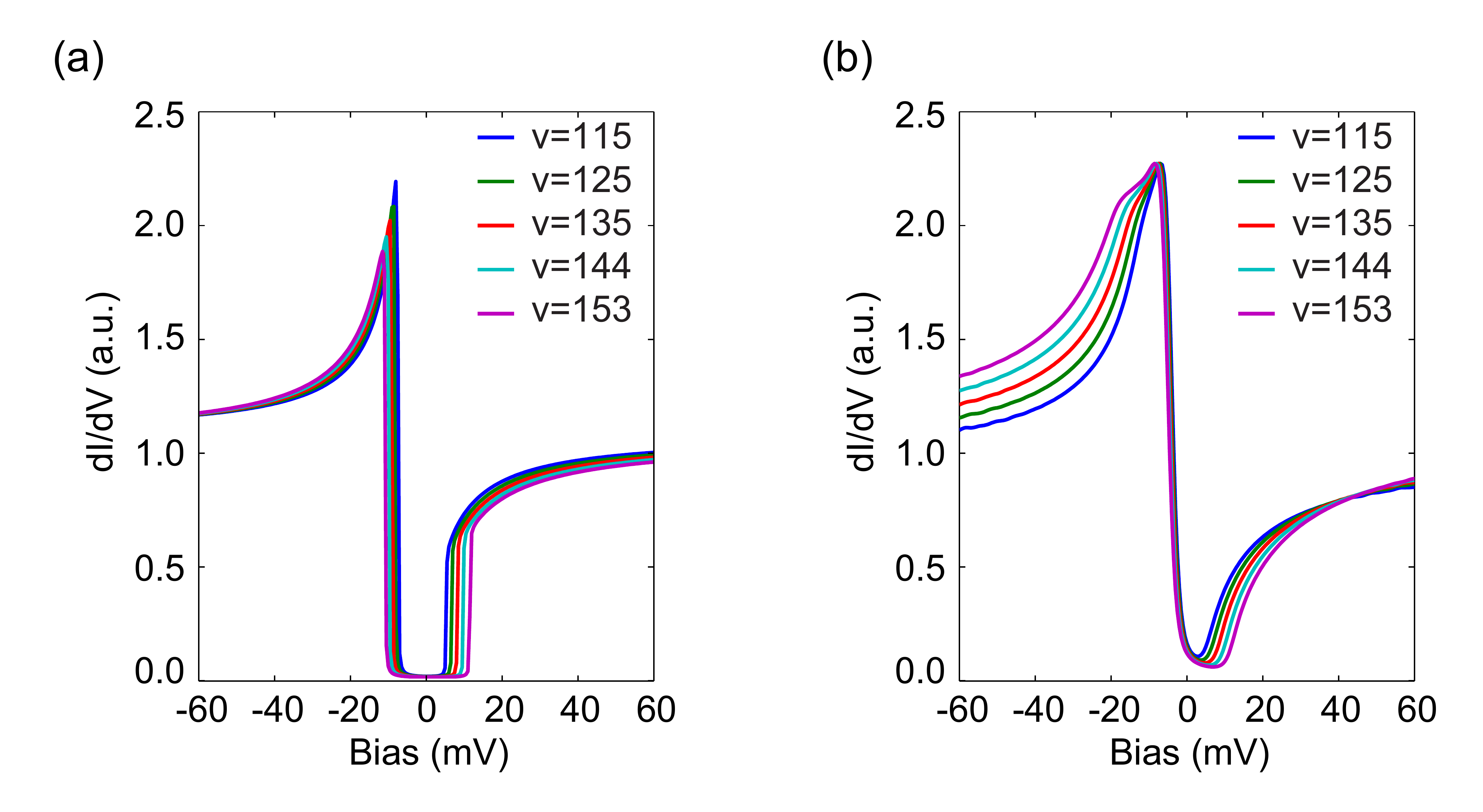}
  \caption{Variation of spectra from Maltseva model (\textbf{a}) and Figgins model (\textbf{b}) with hybridization amplitude $v$. The spectra were calculated using $tf/tc = -0.02$, and the parameters listed in the text.
\label{fig:varyv}
 }
\end{figure*}

We summarize the three models in Figs.\ \ref{fig:varytf} and \ref{fig:varyv}. The calculated $dI/dV$ is plotted as a function of bias and tunneling ratio/Fano parameter. All models agree on the following main features: (1) the relative prominence of a negative energy peak for $t_f/t_c<0$ and the emergence of a positive energy peak for $t_f/t_c>0$; (2) the persistence of a deep gap ($< 10\%$ of the background $dI/dV$) near the Fermi level across a wide range of $t_f/t_c$. However, the Fano model fails to capture some subtleties of the full Kondo lattice models\cite{ZhangPRX2013} which are seen in the data, such as the abrupt kink on the positive edge of the gap.
\\
\\

\phantom{{\Huge O}}\par
\phantom{{\Huge O}}\par
\noindent {\large \textbf{\textsf{References}}}
\vspace{-1.6cm}

%

\end{document}